\documentclass[12pt]{article}

\usepackage{mathptmx}       % selects Times Roman as basic font
\usepackage{helvet}         % selects Helvetica as sans-serif font
\usepackage{courier}        % selects Courier as typewriter font
\usepackage{type1cm}        % activate if the above 3 fonts are
\usepackage{makeidx}         % allows index generation
\usepackage{graphicx}        % standard LaTeX graphics tool
\usepackage{multicol}        % used for the two-column index
\usepackage[bottom]{footmisc}% places footnotes at page bottom
\usepackage{amsmath}

\usepackage[affil-it]{authblk}

\usepackage[font=footnotesize,margin=30pt]{caption}

\usepackage{geometry}
\makeindex             % used for the subject index
\begin{document}

\title{Mental Lexicon Growth Modelling Reveals the Multiplexity of the English Language}
\author{Massimo Stella\thanks{Corresponding author:  \texttt{massimo.stella@inbox.com}.} }

\author{Markus Brede}
\affil{Institute for Complex Systems Simulation, University of Southampton, Southampton, UK} 

\maketitle

\abstract{In this work we extend previous analyses of linguistic networks by adopting  a multi-layer network framework for modelling the human mental lexicon, i.e. an abstract mental repository where words and concepts are stored together with their linguistic patterns.  Across a three-layer linguistic multiplex, we model English words as nodes and  connect them  according to (i) phonological similarities, (ii) synonym relationships and (iii) free word associations. Our main aim is to exploit this multi-layered structure  to explore the influence of phonological and semantic relationships on lexicon assembly over time. We propose  a  model  of lexicon growth which is driven by the phonological layer:  words are suggested according to different orderings of insertion (e.g. shorter word length, highest frequency, semantic multiplex features) and accepted or rejected subject to constraints. We then measure times of network assembly and compare these  to  empirical data about the age of acquisition of words. In agreement with empirical studies in psycholinguistics, our results provide quantitative evidence for the hypothesis that word acquisition is driven by features at multiple levels of organisation within language.}

\section{Introduction}
\label{sec:1}

Human language is a complex system: it relies on a hierarchical, multi-level combination of simple components (i.e. graphemes, phonemes, words, periods) where "each unit is defined by, and only by, its relations with the other ones" \cite{aitchison2012words,baronchelli2013networks}. This definition \cite{aitchison2012words} might explain some of   the success of complex network modelling of language for investigating the cognitive processes behind the so-called human mental lexicon (HML) \cite{baronchelli2013networks}. Psycholinguists  conjecture \cite{aitchison2012words,beckagelanguage,vitevitch2008can} that words and concepts are stored within the human mind in such mental repository, which allows word retrieval according to multiple relationships (i.e. semantic, phonological, etc.). One can imagine the HML as an extensive database, where words are stored together with their linguistic patterns (e.g. synonym relations, etc.) on which a distance metric can be imposed, allowing for comparisons across entries.

In the last fifteen years, different layers of the HML have been investigated using tools from  network theory. Motter et al. \cite{motter2002topology} constructed a semantic network of synonyms, where words appearing as synonyms in a dictionary were connected. The resulting network exhibited small-worldness (i.e. higher clustering coefficient and similar mean shortest path length compared to random graphs). It also displayed a heavy-tailed degree distribution with scaling exponent $\gamma\simeq3.5$. The authors attributed both the presence of network hubs and the small-world feature to \textit{polysemy}, i.e. a given word having more meanings depending on context and thus gathering more links. Sigman and Cecchi \cite{sigman2002global} showed that polysemic links create shortcuts within semantic networks, thus reducing path lengths between semantically distant concepts.  This is relevant to cognitive processing because the semantic topology  correlates with performance in word retrieval in memory tasks \cite{collins1975spreading,de2008word,beckagelanguage,aitchison2012words}. It is conjectured that words within the HML are recollected together with a set of additional properties (e.g. being animated,etc.) \cite{collins1975spreading}. Empirical evidence supports the hypothesis that adjacent words in a semantic network inherit features from their neighborhood, so that words closer on the network topology can be processed in a correlated way, thus reducing memory effort \cite{beckagelanguage,aitchison2012words,collins1975spreading}. Semantic networks were further analysed by Steyvers and Tenenbaum in \cite{steyvers2005large}. By proposing a network growth model based on preferential attachment, the authors investigated the role of word learning variables (e.g. frequency and age of acquisition) on shaping the structure of semantic networks. They showed that higher frequency words tend to have more semantic connections and tend to be acquired at earlier stages of development , thus highlighting an interplay between network topology and language learning.  

Complex networks were also proposed as a suitable tool for analysing the phonological layer of the HML. Vitevitch suggested phonological networks (PNs) \cite{vitevitch2008can} as complex networks in which  words  are connected if phonologically similar, i.e. if they differ by the addition, substitution or deletion of one phoneme. Experimental evidence showed that the resulting network degree and local clustering coefficient both correlated positively with speech errors and word identification times, indicating that also the topological properties of a word in the phonological network  plays an important role in  its cognitive processing \cite{vitevitch2014insights}. In \cite{Stella2014,stella2015markov}, we checked that artificial corpora, made of uncorrelated random words, could not reproduce specific features of the English phonological network. By means of percolation experiments we showed that the real PN actually inherits some features (e.g. a degree distribution with a heavy tail) from its embedding space, but it also displays some patterns that are extremely hard to match with random word models (e.g. the PN's empirical core-periphery structure).
By proposing a family of  null models that respect the spatial embedding,  we identified two constraints possibly acting on phoneme organisation: (i) a maximum size of phonological neighborhoods (above which word confusability \cite{vitevitch2014insights,vitevitch2008can} becomes predominant) and (ii) a tendency to avoid local clustering  (which correlates with word confusability \cite{vitevitch2014insights}).

To the best of our knowledge, until now there has been no theoretical framework modelling both the semantic and the phonological aspects of the mental lexicon in terms of a multiplex network. Multiplexes represent a novel and quite prolific research field \cite{battiston2014structural,boccaletti2014structure}: in a multiplex the same set of nodes can be connected differently in different layers of networks. Historically, the idea of context-dependent links originated from the social sciences \cite{battiston2014structural}. However, it is only in the last five years that these multi-layered networks were successfully applied in a wide collection of different scenarios, such as robustness of infrastructure, science of science and game-theoretic dilemmas, among many others (for further references see \cite{boccaletti2014structure}).

Exploring the multiplexity of the English language to study lexicon formation is the main idea of this study. We specifically focus on the interplay between semantics and phonological factors in the assembly of the repertoire over time.  In detail, we build a three-layer multiplex network, where each layer represents a given linguistic network and where the same set of nodes is replicated across all layers. We focus our analysis on a minimalistic network growth model where the lexicon is assembled over time and real words get inserted, one at a time, according to a given ordering, either based on exogenous features (e.g. word frequency) or  multiplex features of the HML. Our main aim is to quantify the influence of each ordering in the assembly times of the empirical multiplex, in order to assess the impact of word features on  lexicon growth. For this purpose, we test our experiments with empirical data of the age of acquisition of  English words, obtained from \cite{kuperman2012age}. Our  results highlight the  presence of an interplay between the phonological and the semantic layers in structuring the mental lexicon. 

This paper is structured as following: in Sect. \ref{sec:2} we report on the dataset we adopted for the multiplex construction and we compare it to datasets of commonly spoken English; in Sect. \ref{sec:3} we introduce the model of lexicon growth; the results are discussed in Sect. \ref{sec:4}, conclusions and future work directions are reported at the end.

\section{Multiplex Construction}
\label{sec:2}
We build a linguistic multiplex of three unweighted, undirected graphs/layers, comprising an intersection of $N=4731$ words and based on the following interactions: 
\begin{enumerate}
\item{Free Word Associations (based on the Edinburgh Associative Thesaurus \cite{kiss1972associative});}
\item{Synonyms (based on WordNet 3.0 \cite{miller1998wordnet});}
\item{Phonological Relationships (based on WordNet 3.0 and manually checked, automatic phonological transcriptions into the IPA alphabet \cite{bronstedautomatic}).}
\end{enumerate}

The thus constructed linguistic multiplex includes one phonological layer and two semantic layers. With synonym relationships and word associations we chose to include two semantic layers, mainly because of large structural differences in the topology of these networks. Free associations capture also those linguistic patterns that cannot be expressed in terms of other semantic relationships (e.g. opposites, synonyms, etc.). These relationships are still of primary importance for  cognitive processes \cite{kiss1972associative,beckagelanguage}. In fact, experimental evidence indicates that such links act as pointers for word retrieval \cite{aitchison2012words,beckagelanguage}. Their greater generality is what differentiates associations from synonymy relationships, which have been extensively investigated in the linguistic literature \cite{steyvers2005large,beckagelanguage,de2008word}.

\subparagraph{Representativeness of the Data} %
A network representation of language should be indicative of real patterns in the mental lexicon, therefore the linguistic multiplex should be based on commonly used words. Unfortunately WordNet 3.0 does not include frequency counts, therefore we tested our data through the word frequencies from the Opensubtitles dataset \cite{barbaresi2014language}, i.e. a lexicon based on more than $1.4\cdot10^{8}$ word counts from TV series subtitles. 

\begin{figure}[t]
\centering
% For example, with the graphicx style use
\includegraphics[scale=.42]{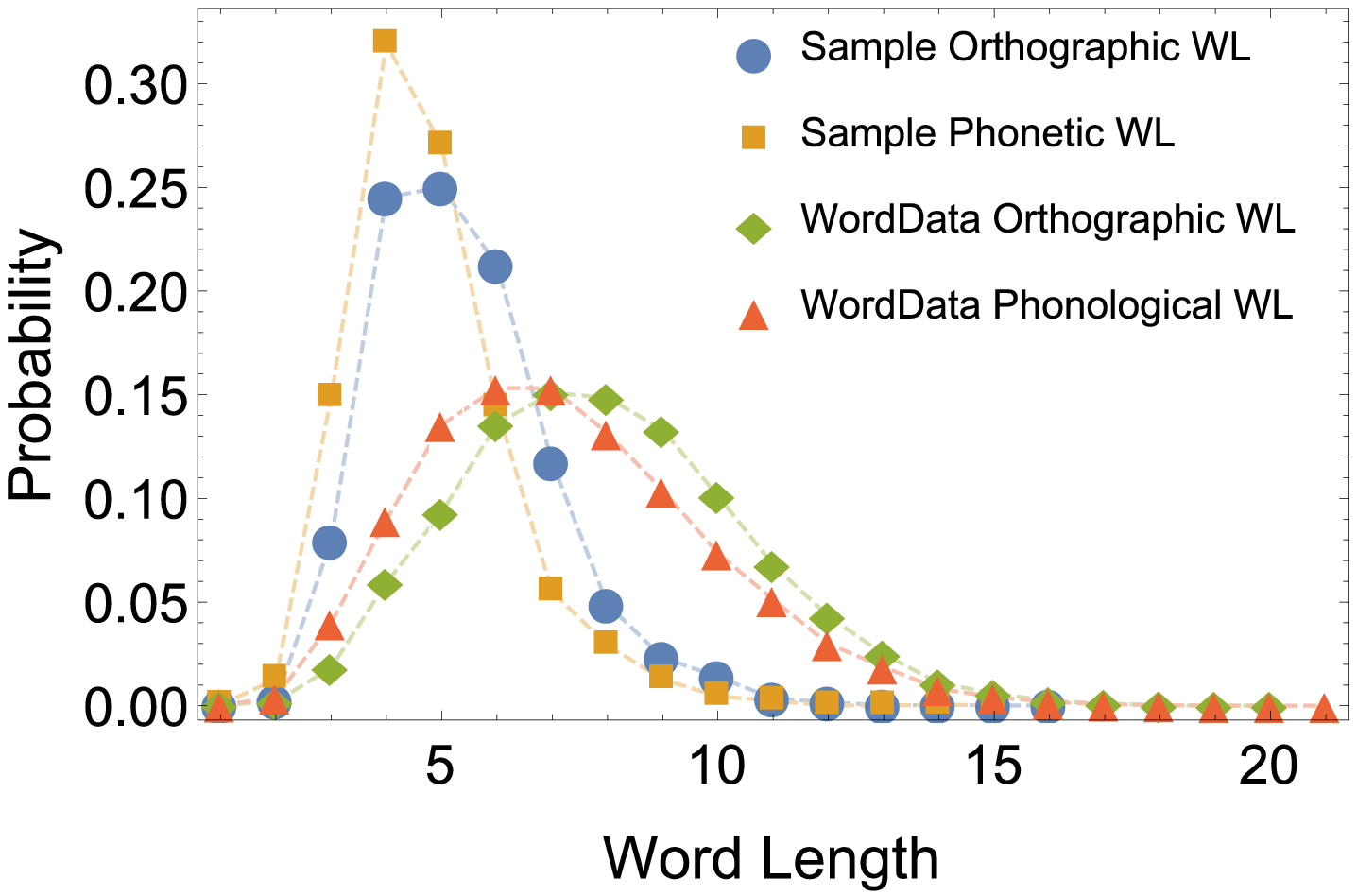}
\includegraphics[scale=.42]{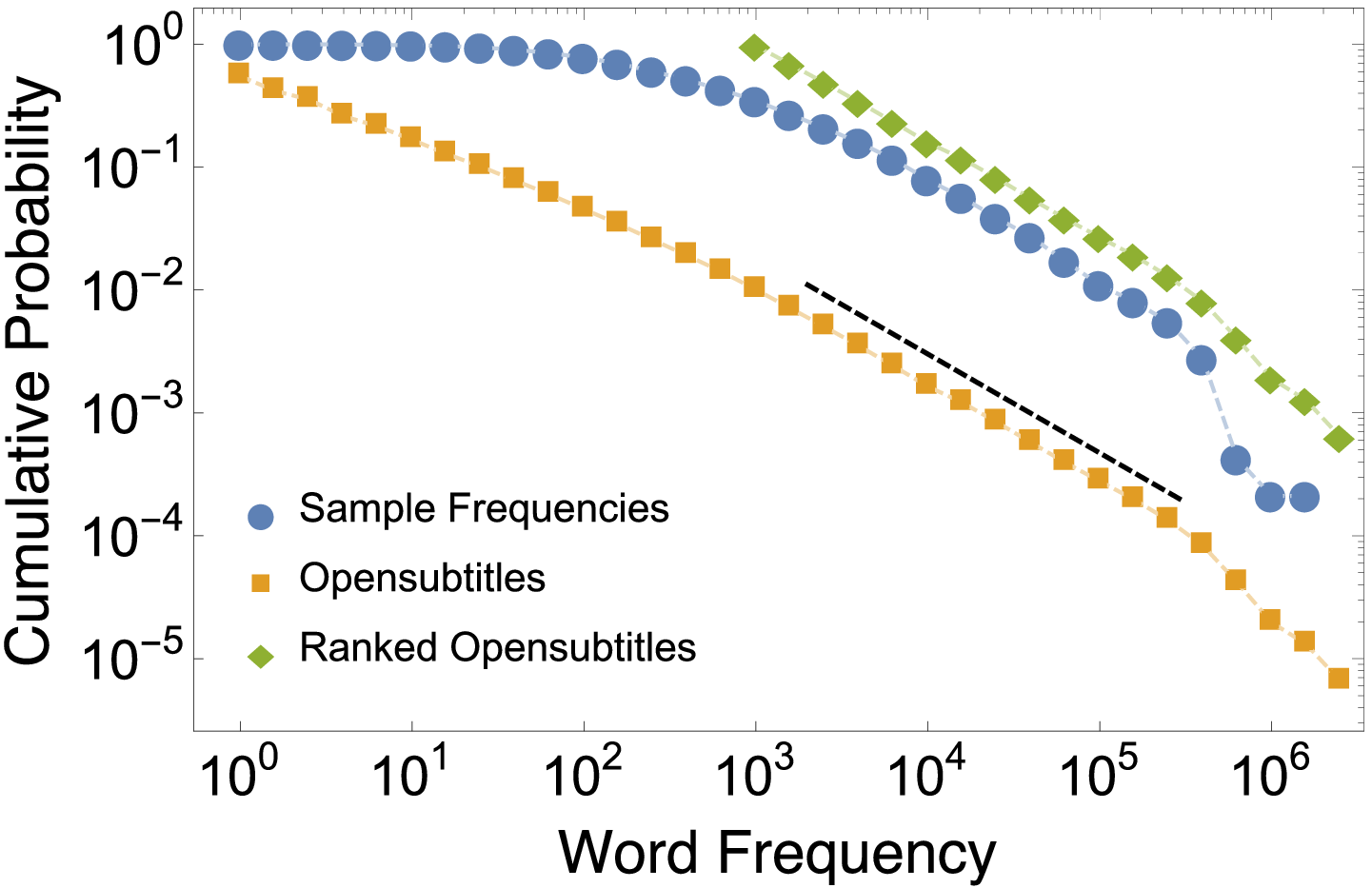}
%
% If no graphics program available, insert a blank space i.e. use
%\picplace{5cm}{2cm} % Give the correct figure height and width in cm
%
\caption{Left: Orthographic and phonetic word length distributions for our sample (blue dots and golden squares, respectively) and for roughly 29000 words phonetically transcribed within WordNet 3.0 (green diamonds and red triangles). Opensubtitles is not used in the word length distributions because it does not have phonological transcriptions. Right: Empirical probability distributions of word frequency within our data sample (blue dots), the Opensubtitles repository (golden squares) and a ranked subsample of the Opensubtitles list of the same size as our  sample (green diamonds). The dashed black line gives a power-law with exponent $\gamma=1.83\pm0.03$.}  % Give a unique label
\label{fig:1} 
\end{figure}

The word length distributions reported in Fig.~\ref{fig:1} indicate that our smaller-size word sample contains more shorter words when compared to WordNet. Furthermore, Fig.~\ref{fig:1} shows that words contain less phonemes than orthographic characters, on average. For instance, a word in our sample contained $4.78\pm0.03$ phonemes and $5.39\pm0.03$ orthographic characters. Given this difference, we are using both phonetic and orthographic word lengths in our growth experiments in Sect. \ref{sec:4}, as proxies for word acquisition through hearing and reading, respectively. 

In Fig. ~\ref{fig:1} the word frequency distribution of our 4731 sampled words is compared against the whole Opensubtitles repository and against the 4731 words from Opensubtitles with the highest frequencies. Interestingly, the whole dataset exhibits a heavy tail behaviour. The cumulative probabilities $P(F\geq z)$ of finding a word with frequency $F$ greater than or equal to $z$ tell us that higher frequency words in our dataset are more likely than in the whole Opensubtitles but also less likely than in the frequency ranked subsample. Furthermore, excluding extremely frequent words, our sample reproduces the same power-law like behaviour of the whole Opensubtitles dataset, for mid-and high frequencies.  Because of this over-representation of higher frequency words, we can reasonably assume that our data is a good representation of commonly used English words

\subparagraph{Multiplex Network Structure} %

\begin{figure}[t]
\centering
% For example, with the graphicx style use
\includegraphics[scale=.61]{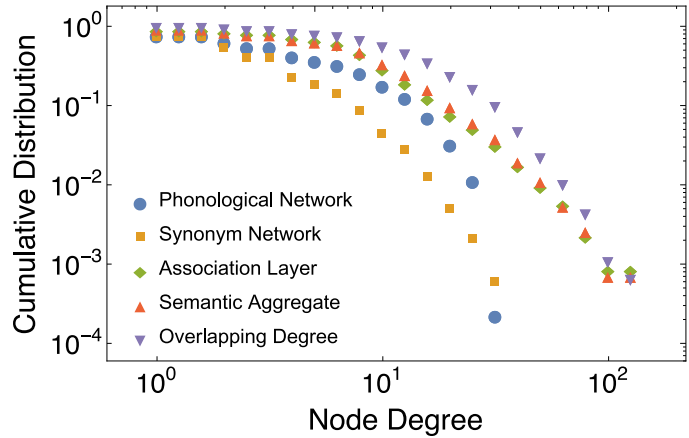}
%
% If no graphics program available, insert a blank space i.e. use
%\picplace{5cm}{2cm} % Give the correct figure height and width in cm
%
\caption{Cumulative degree distribution $P(K\geq k)$ for the phonological network (blue dots), the synonym network (golden squares), the free association network (green diamonds), the semantic aggregate network associations+synonyms (red triangles) and the overlapping network (purple triangles).}
\label{fig:1a} % Give a unique label
\end{figure}

We begin the analysis of the multiplex by investigating  the cumulative degree distributions $P(K\geq k)$ \cite{newman2010networks} of individual layers and of the multiplex, reported in Fig.~\ref{fig:1a}. The degree distributions span different orders of magnitudes and display different behaviors. There is a considerable fraction of hubs within the association network, which displays a heavy tail degree distribution. The phonological network displays a cut-off around degree $k \approx 30$ while the synonym network shows a degree distribution that can be approximated by an exponential.  We also investigate the multiplex \textit{overlapping degree} $o_{i}$  \cite{battiston2014structural}, i.e. the sum of degrees $k_{i}^{[\alpha]}$ of node $i$ on each layer $\alpha$:
\begin{equation} 
\label{ovl_eqn} 
o_{i}=\sum_{\alpha=1}^{3}k_{i}^{[\alpha]}.
\end{equation}

Interestingly, the overlapping degree seems to have a more pronounced exponential decay compared to the degree of the \textit{aggregated semantic layers} (i.e. a network where any link is present if it is present in at least one of the original layers). This reveals negative degree correlations on lower degrees, in our linguistic multiplex. A scatter plot highlights the presence of hub nodes in the semantic aggregate that have low degrees ($k \leq 15$) in the phonological layer. For this reason, locally combining layer topologies is of interest for our assembly experiments, cf. Sect. \ref{sec:4}.

Table \ref{tab:1} reports some network metrics \cite{newman2010networks} for the individual layers and the aggregated semantic multiplex, compared to configuration models with the same degree distributions. All the three layers display the small-world feature, in agreement with previous results  \cite{motter2002topology,Stella2014}. The current literature suggests that small-worldness might be related to language robustness to individual word retrieval failure (e.g. in aphasia \cite{beckagelanguage}) while  also enhancing network navigability \cite{vitevitch2014insights}. It is noteworthy that the phonological layer displays a network diameter almost three times larger than the mean path distance. Since its configuration model (CM) counterpart does not reproduce such pattern, this is an indication of a strong core-periphery structure within the network \cite{Stella2014}. Further, all the individual layers are disconnected and have a giant component (GC). The GC size is hardly matched by CMs for the phonological and synonym layers while there is good agreement for the association layer. Interestingly, the two semantic layers display different organisational features: the synonym layer is more disconnected but more clustered than the association one; and while synonyms display an assortative mixing by degree associations are disassortative \cite{newman2010networks}, instead. Therefore, in the association layer there are hub words surrounded by many poorly connected nodes while in the synonym layer large neighborhoods tend to be directly connected with each other. Indeed, because of these different topological features we will keep these layers distinct within our linguistic multiplex. Assortative mixing is also strongly present at the phonological level, but note that in this case  high assortativity is a feature inherited from the embedding space of phonological networks \cite{Stella2014,stella2015markov}. Configuration model aggregates are formed by aggregating the individual configuration model layers. Only 0.5\% of the edges in the association layer overlap with the synonym layer. The empirical networks display a higher edge overlap when compared to the configuration models ($4.5\%$ of the edges overlap across the real layers versus the $0.1\%$ of the CMs).

\begin{table}
\footnotesize
\caption{Metrics for the multiplex with $N=4731$ nodes: edge count $L$, average degree $\left\langle k\right\rangle$, mean clustering coefficient $CC$, assortativity coefficient $a$, giant component node count $GC Size$, network diameter $D$ and mean shortest path length $\left\langle d\right\rangle$. Error bars on the last digit are reported in parentheses and are based on 20  repetitions. For instance, $3.0(4)$ means $3.0\pm0.4$. CM aggregates are obtained by combining CM layers and therefore differ in degree from their empirical counterparts.}
\label{tab:1} % Give a unique label
\begin{tabular}{p{4cm}p{1.cm}p{0.75cm}p{1.3cm}p{1.5cm}p{1.3cm}p{0.3cm}p{0.6cm}}
\hline\noalign{\smallskip}
Network & $L$ & $\left\langle k\right\rangle$ & $CC$ & $a$ & $GC Size$ & $D$ & $\left\langle d\right\rangle$ \\
\hline\noalign{\smallskip}
\rule{0pt}{1em}Phonological & 15447 & 6.5 & 0.24 & 0.61 & 3668 & 22 & 6.7  \\
\rule{0pt}{1em}Phonological CM & 15447 & 6.5 & 0.004(1) & 0.0048(4) & 4580(10) & 10 & 4.3(5) \\
\rule{0pt}{1em}Synonym & 7010 & 3.0 & 0.23 & 0.26 & 2989 & 15 & 5.9 \\
\rule{0pt}{1em}Synonym CM & 7010 & 3.0 & 0.002(1) & -0.02(3) & 3396(9) & 13 & 5.1(2) \\
\rule{0pt}{1em}Association & 20375 & 8.6 & 0.1 & -0.11 & 3664 & 7 & 3.6 \\
\rule{0pt}{1em}Association CM & 20375 & 8.6 & 0.09(2) & -0.005(1) & 3658(8) & 7 & 3.5(3) \\
\rule{0pt}{1em}Semantic Aggregate & 26056 & 11. & 0.18 & -0.06 & 4298 & 9 & 3.6 \\
\rule{0pt}{1em}Sem. Agg. Combined CMs & 27374 & 11.6 & 0.01(2) & 0.015(1) & 4320(10) & 8 & 3.5(2) \\
\rule{0pt}{1em}Multiplex Aggregate & 40983 & 17.3 & 0.15 & 0.018 & 4689 & 9 & 3.4 \\
\rule{0pt}{1em}Mult. Agg. Combined CMs & 42787 & 18.1 & 0.012(5) & 0.024(4) & 4713(6) & 8 & 3.2(1) \\
\noalign{\smallskip}\hline\noalign{\smallskip}
\end{tabular}
\end{table}

\section{Simulated Network Assembly}
\label{sec:3}

In \cite{Stella2014}, we suggested a network  growth procedure as a null model for phonological networks (PNs), in which an artificial PN was built from randomly assembled strings of phonemes, satisfying some empirical constraints (e.g. phoneme frequencies). In this work, we extend that model by adopting a multiplex perspective. 

Let us model the mental lexicon as a network which grows over time. Our model is \textit{localist} \cite{steyvers2005large}, i.e. in it each concept is partially associated with an individual node/word in the network. Concepts are acquired to the lexicon by inserting single nodes/words. However, a given concept is represented in its full meaning by a word/node together with its links, since they retain further information about the concept itself (e.g. a neighborood can translate into a semantic context \cite{steyvers2005large} or it can provide information about  word confusability \cite{vitevitch2008can}). In the following we will use "words" to identify single nodes and "concepts" to identify jointly a node and its local connectivity. We follow an approach similar to Steyvers and Tenenbaum \cite{steyvers2005large}.

At each time step, a node/word is tentatively inserted into the lexicon. Then, we check for phonological similarities between the new word and the others already in the network, i.e. we check for links on the phonological level. If the new node/word receives at least one connection (i.e. it becomes \textit{active} in the multiplex jargon \cite{boccaletti2014structure}) on the phonological network, then it is accepted to the lexicon. Otherwise, if the node/word does not receive any connection, we reject it  with probability $f$, putting it back to the list of not yet included words. Words are suggested from this list according to a given multiplex or exogenous criterion and until all words have been accepted. We measure the average assembly time $T$, i.e. the time it takes until a full network comprising all 4731 words has been built. The rejection probability $f$ is the only free parameter of the model, but acceptance/rejection of words also depends strongly on the ordering in which they are suggested. There are many possibilities of different orderings that could be considered. We tested several of them and then selected a sample of those experiments that provided a wide pool of different results:
\begin{enumerate}
\item{random ordering as a baseline reference case (Rand. Order);}
\item{phonologically shorter words first (Short Pho., e.g. "a", "ad", "ash", ...) ;}
\item{orthographically shorter words first (Short Wor., e.g. "a", "ad", "be", ...) ;}
\item{more frequent words first (Freq., e.g. "a", "in", "have", ...);}
\item{higher degree words in the association layer first, where hubs are the most recollected words in semantic memory (Asso., e.g. "man", "water", "sex", ...);}
\item{higher degree words in the synonym layer first, notice the difference with the association layer in the ranking (Syno., e.g. "take", "hold", "get", ...);}
\item{higher degree words in the semantic multiplex aggregate first, association hubs prevail over the synonyms (As.+Sy., e.g. "man", "water", "sex", ...);}
\item{empirical age of acquisition \cite{kuperman2012age} (AoA, e.g. "momma", "potty", "water", ...);}
\item{random phonological/random semantic neighbors, i.e. select a word at random on the phonological level, select one of its neighbors on the semantic aggregate at random, avoiding repetitions (R. Ph./Ag.);}
\item{random phonological/frequent semantic neighbors, i.e. select a phonological word at random, select one of its neighbors on the semantic aggregate at random but proportionally to its frequency (R. Ph./F. Ag.);}
\item{frequent phonological/frequent semantic neighbors, i.e. select a phonological word at random but proportionally to its frequency, similarly select one of its neighbors on the semantic aggregate (F. Ph./Ag.).}
\end{enumerate}

In our model the growth dynamics is driven by the phonological layer. Although this could be made more realistic, our choice is motivated by two  empirical observations. Firstly, there is  widely accepted empirical evidence showing that phonological memory (i.e. the growing set of phonological transcriptions that are checked for connections, in our model) plays a critical role in concept acquisition   \cite{ferguson1975words,schwartz1982children,stoel1984patterns,hoff2008non,wiethan2014early}. Furthermore, there are recent empirical studies in children that  strongly emphasise  that lexical acquisition is heavily influenced by the phonology of the words, at least at early stages of the lexicon's assembly \cite{wiethan2014early}. Psycholinguists conjecture that this lead of phonology in the lexicon growth might occur because children could find it easier to produce and understand words containing phonemes already presented in their phonological inventory \cite{schwartz1982children,aitchison2012words}. This empirical bias is what our model tries to capture by checking for phonological similarities before word acceptance/rejection. However, it is also true that semantics and other external features do influence the lexicon's growth and our model does  account for this  interplay through the orderings of word insertion. In fact, there is also evidence that, after an initial state in which phonological learning is predominant, lexical learning lets children learn novel words whose sounds are not present in their inventories \cite{wiethan2014early,hoff2008non}. Our model captures also this aspect, since even novel words that do not have phonological similarities can be probabilistically accepted. The second motivation behind adopting the phonological network as a check for linguistic relationships is that detecting phonological similarities is straightforward: it can be done on a quantitative basis (i.e. check for phoneme strings having edit distance one). Conversely, detecting semantic relationships (i.e. are two words synonyms?) can be extremely difficult without any external source of information (e.g. a dictionary or an experiment). 

Beyond the type of links we check, another key element of our model is the "activation" requirement, i.e. the fact that a word has to receive at least one connection in order not to undergo the probabilistic rejection/acceptance stage. Being connected to any other node is the simplest requirement one can think of in terms of local connectivity, which is pivotal in the activation spread \cite{collins1975spreading}. We have made this modelling choice mainly in the interests of meaningful parsimony. While we do not mean to preclude a possible role for other growth dynamics, we have to start from a simple, yet meaningful, dynamics that minimises the number of free parameters. It has to be underlined that our chosen model represents, at best, a highly simplified abstraction of the cognitive processes driving real lexicon growth. We chose to follow simplicity, mainly because of how little is known about the evolution of the real, large-scale human mental lexicon \cite{steyvers2005large,aitchison2012words}. Other viable approaches that might fall in the same simplicity category as our model should also be explored in the future.  

Interestingly, the same word is used in both the multiplex and the psycholinguistics jargons: an "active" node in a multiplex is one having at least one link \cite{boccaletti2014structure}, the "activation" in psycholinguistics is a theoretical stimulus signal that spreads through connections across the semantic and/or phonological layers of the mental lexicon when words are to be identified and retrieved \cite{collins1975spreading,beckagelanguage,vitevitch2014insights}. Indeed, our model accepts preferentially "active" (in the multiplex jargon) and potentially "activable" words (in an activation spread model scenario). Our focus on local connectivity was inspired by previous models of lexicon growth \cite{steyvers2005large,beckage2015semnet}, which conjecture that memory search processes might be sensitive to the local connectivity of concepts.

\section{Results and Discussion}
\label{sec:4}

\begin{figure}[t]
\centering

% For example, with the graphicx style use
\includegraphics[scale=.63]{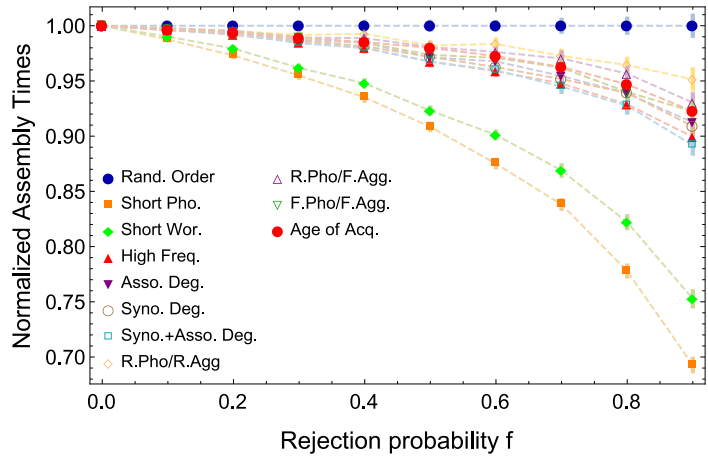}
%
% If no graphics program available, insert a blank space i.e. use
%\picplace{5cm}{2cm} % Give the correct figure height and width in cm
%
\caption{Normalized assembly times for different orderings at different rejection probabilities $f$. These normalized times indicate the average time necessary for the network to get assembled through a given ordering rescaled to the random reference case. Error bars indicate standard errors and are evaluated over 20 different runs.}
\label{fig:3}       % Give a unique label
\end{figure}

In Fig.~\ref{fig:3} we report the normalized assembly times $T/ T_{random}$ for  different orderings for several values of the rejection parameter $f$. These values are rescaled to the random case. Interestingly, such rescaling shows that inserting words with our selected orderings always decreases the assembly times compared to the random scenario. This effect becomes more evident when the rejection probability $f$ is large. For instance, when the probability of rejecting each inactive word is $f=0.8$, inserting words ordered according to their phonetic length (shorter first) fully assembles the network in roughly 78\% of the time necessary in the random case and this distinction is statistically significant. Notice that a-priori, we chose orderings loosely inspired by a least memory effort principle  \cite{aitchison2012words} so that this general trend is expected. Nonetheless, there is an interesting variety of behaviors that need further analysis.

Intuitively, inserting shorter phonetic-length words first is the optimal case in terms of minimum assembly time. Orthographic word length gives slightly higher assembly times. Inserting words according to their frequency gives results that are very close to semantic measures such as the degree rankings in the semantic layers/aggregate and to multiplex features. All these orderings show a trend close to the one where words are inserted within the growing lexicon according to their age of acquisition. Since assembly times are the quantitative proxies of our model for the likelihood of the mechanisms underlying lexicon growth, we adopt the times based on the age of acquisition as another reference point for testing the influence of the other orderings.

\begin{figure}[t]
\centering
% For example, with the graphicx style use
\includegraphics[scale=0.48]{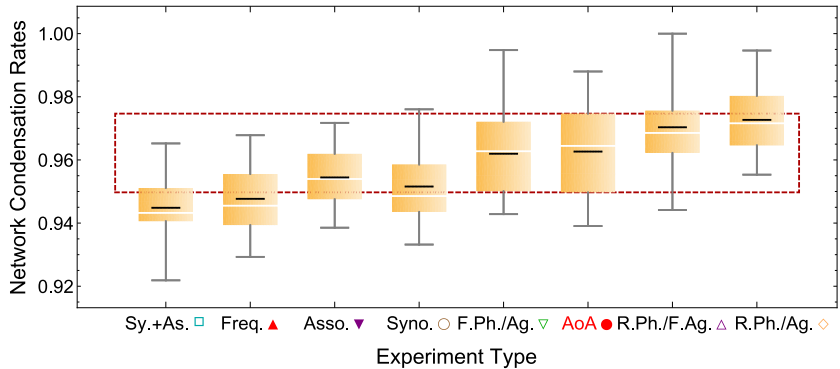}
%
% If no graphics program available, insert a blank space i.e. use
%\picplace{5cm}{2cm} % Give the correct figure height and width in cm
%
\caption{Normalized assembly times of different orderings for a rejection probability equal to 0.8. The age of acquisition is highlighted in red. Whiskers represent distribution extremes while interquartile ranges are represented by orange boxes. White dashes indicate medians while black dashes represent means, instead. Interquartile overlaps represent the fraction of orange boxes falling within the ranges of the age of acquisition scenario.}
\label{fig:4}       % Give a unique label
\end{figure}

We start from the distributions of the assembly time for each ordering, at several values of the rejection probability $f$. We then quantify the overlap of the interquartile range of the age of acquisition case with the other scenarios. We consider the overlap of interquartile ranges rather than the overlap of the whole distributions because interquartile ranges represent a more robust measure of scale against fluctuations on extreme values in small, skewed empirical distributions as ours \cite{moore1989introduction}. Also, interquartile ranges are easy to compute and visualise by commonly used box plots \cite{moore1989introduction}. An example is reported in Fig.~\ref{fig:4}, where a box plot for the interquartile ranges of all our orderings are reported for $f=0.8$. For instance, in that case the frequency ordering does not give results compatible with the empirical case (even though it is very close to the ordering with the degree in the semantic aggregate). Further, considering only the semantic degrees gives a slightly stronger overlap, but is not yet compatible with the age of acquisition case.  Ordering words by their phonological and the semantic network degrees  gives the closest results to the empirical age of acquisition scenario. We interpret this result as a quantitative proof of the importance of the multiplex structure of human language in shaping organisational features of the human mental lexicon. Locally navigating across the linguistic multiplex with a word frequency bias gives the best, highest overlapping results, within the framework of our theoretical model.

\begin{figure}[t]
\centering

% For example, with the graphicx style use
\includegraphics[scale=.735]{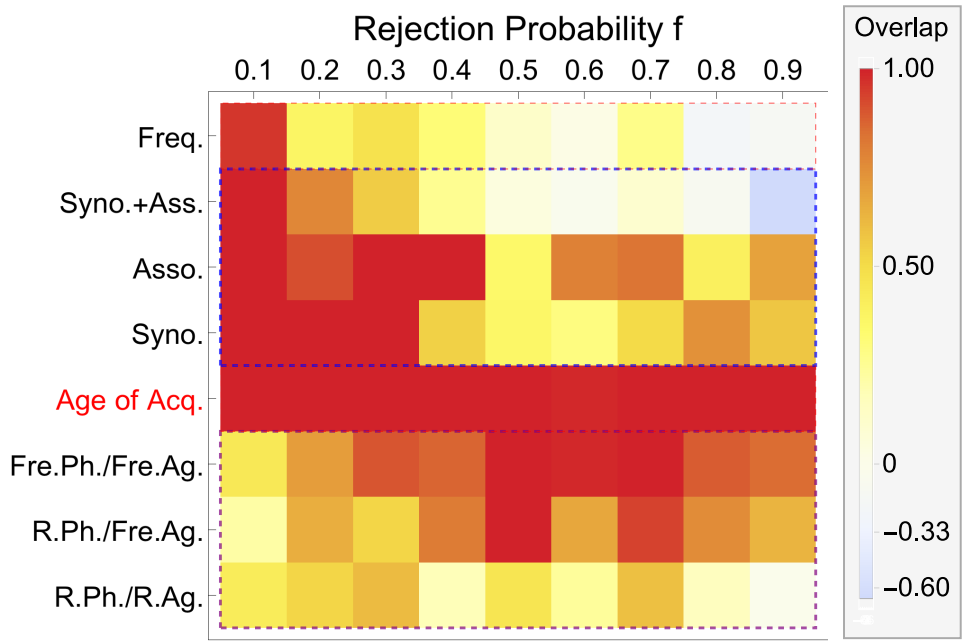}
%
% If no graphics program available, insert a blank space i.e. use
%\picplace{5cm}{2cm} % Give the correct figure height and width in cm
%
\caption{Heat map of the normalized overlaps of the interquartile ranges of the assembly times relative to the age of acquisition case. The colours indicate: red = a perfect overlap (see the age of acquisition row); white = the absence of overlap; blue = the respective interquartile ranges are quite far. Orderings based on multiplex features are highlighted, the semantic ones on top and those based on multiplex neighborhoods on the bottom.}
\label{fig:5}       % Give a unique label
\end{figure}

In Fig.~\ref{fig:5} we checked the performance of the multiplex-based ordering versus $f$. Let us underline that during a given assembly $f$ is kept fixed. However, when the probability of rejecting unconnected words is low, the orderings based only on either frequency or the semantic degrees perform relatively well. We can think of this stage as the real lexical learning phase \cite{wiethan2014early,beckagelanguage,beckage2015semnet,stoel1984patterns}, which happens later in language development and where novel words are inserted within the lexicon according to their semantic information and almost independent of phonological similarities. Larger values of the rejection probability $f$ correspond to scenarios where the frequency and semantic degree orderings give results significantly different from the age of acquisition case. We can interpret this stage as a phonological learning phase \cite{ferguson1975words,schwartz1982children,stoel1984patterns}, where words are inserted to the lexicon strongly based on their phonological similarities and where the phonological and the semantic layers are strongly interdependent. Therefore, our model highlights an interesting shift from one strongly semantic to a strongly multiplex stage, depending on the $f$ parameter. This is a first quantitative finding about the importance of a multiplex modelling of the human mental lexicon. In fact, partial knowledge as frequency or phonological information only is unable to reproduce the same patterns across the whole parameter space.

\section{Conclusions and Future Work}
\label{sec:5}

Here we proposed  a simplified model of lexicon growth which is based on  a representation of the English HML on several levels via the framework of multiplex networks. Motivated by empirical evidence and technical advantages in checking for phonological links we focus on the phonological level for the growth dynamics. 

Numerically estimated assembly times identify a higher likelihood of a lexicon growth encapsulating information from the multiplex structure of free associations, synonyms and phonologically similar words, compared to assembly based on information from single layers or only word frequencies or lengths. In fact, assembly times can be thought of as proxies for the likelihood of the mechanisms underlying lexicon growth. When words are acquired without strong phonological biases (as in later stages of children's linguistic development) then orderings based on frequency and on semantic local centralities (i.e. node degree) are in good agreement with the empirical case. On the other hand, when words are acquired with stronger biases, as it happens in earlier stages of children's linguistic development, orderings based on the multiplexity of the English language provide results closer to the real scenario. 

There are many interesting questions that this preliminary work opens. The first is a more extensive investigation of the multiplex features of the English language, e.g. a more detailed structural investigation of multiplex reducibility, layer overlap, cartography, clustering, efficiency and robustness to word retrieval failure \cite{boccaletti2014structure}. Another interesting research direction would be trying to generalize our model by basing acceptance on the formation of more than one connections, or rather on links created also on other multiplex layers different from the phonological one, possibly by using the empirical semantic connections as a reference. This generalisation would be more realistic but also more cumbersome in adding more parameters to a model, which, already in this simple version, is capable of displaying an interplay between lexical and phonological learning. 

From a complex systems perspective, it would be interesting to explore further the "multiplexity" of the English language, namely the interplay between phonological and semantic features, also by comparing the model against real data from children. Last but not least, a multiplex analysis for languages different from English could represent an interesting theoretical framework for testing both distinctive and universal features of human language.

\section{Acknowledgements}
MS acknowledges the DTC in Complex Systems Simulation, University of Southampton for financial support. The authors acknowledge Dr. Srinandan Dasmahapatra, Nicole Beckage and the reviewers for providing insightful comments.

\bibliographystyle{plain}
\bibliography{biblu3}
\end{document}